\documentclass[a4paper]{jpconf}
\usepackage{graphicx}
\begin{document}
\pagenumbering{arabic}
\pagestyle{plain}

\title{A Second Detector Focusing on the Second Oscillation
Maximum at an Off-axis Location to Enhance the Mass Hierarchy 
Discovery Potential in LBNE10}
\author{X. Qian$^{a,b}$, J. J. Ling$^{a}$, R. D. McKeown$^c$, W. Wang$^c$, E. Worcester$^a$, C. Zhang$^a$}
\address{$^a$Physics Department, Brookhaven National Laboratory, Upton, NY}
\address{$^b$Kellogg Radiation Laboratory, California Institute of Technology, Pasadena, CA}
\address{$^c$College of William and Mary, Williamsburg, VA}
\ead{xqian@bnl.gov}

\begin{abstract}
The Long Baseline Neutrino Oscillation Experiment (LBNE) is proposed to determine 
the neutrino mass hierarchy and measure the CP phase $\delta_{CP}$ in the leptonic 
sector. The current design of LBNE Phase I consists of a 10 kt liquid argon time projection chamber 
(LBNE10). The neutrino-antineutrino asymmetry in the electron-neutrino appearance probability 
has contributions from both the CP phase and the matter effect. For this reason, experimental 
sensitivity to the mass hierarchy depends both on the true value of the CP phase and the true 
mass hierarchy; LBNE10 will determine the mass hierarchy at high levels of significance for 
half of $\delta_{CP}$ phase space.
%Due to the potential enhancement or cancellation between the matter 
%effect (depending on the neutrino mass hierarchy) and the CP effect (depending
%on the value of CP phase $\delta_{CP}$), the chance for the mass hierarchy determination 
%is excellent in one half of the $\delta_{CP}$ phase space (the good half of $\delta_{CP}$), 
%but becomes  marginal in the other half of the $\delta_{CP}$ phase space (the bad half 
%of $\delta_{CP}$). 
We propose placing a second detector at an off-axis location. 
Such a detector will share the same beamline as the primary LBNE detector. 
The detector location is chosen such that this detector 
focuses on a measurement of electron (anti-)neutrino appearance at the 
second oscillation maximum. 
%We will show that this configuration will 
%enhance the mass hierarchy discovery potential in LBNE10 and  strengthen the measurement 
%of $\delta_{CP}$ in the leptonic sector.
We will show that this configuration will enhance the ability of LBNE to determine the mass 
hierarchy and to discover CP violation in the leptonic sector.
\end{abstract}

The recent discovery of sizable $\theta_{13}$ with reactor electron anti-neutrino 
disappearance measurements~\cite{Dayabay_1, reno, doublec, Dayabay_2} and electron 
(anti-)neutrino appearance measurements~\cite{t2k, Minos} opens door to the 
determination of the neutrino mass hierarchy (MH) and the CP phase $\delta_{CP}$ 
in the leptonic sector~\cite{bob_review}.  The neutrino mass hierarchy
problem is to determine whether the
third generation of neutrino is heavier (normal hierarchy/NH) or lighter (inverted 
hierarchy/IH) than the first two generations of neutrinos. 
Together with 
the next generation neutrinoless double beta decay experiments, the determination
of the neutrino mass hierarchy may hold the key to the nature of neutrinos 
(Dirac or Majorana particles). A value of the CP phase not equal to zero or $\pi$ may 
produce the CP-violation that is required for leptogenesis~\cite{lepto}, which is a 
potential explanation for the apparent matter-anti-matter asymmetry in the universe.
%A non-zero CP phase 
%(also non-$\pi$) could strengthen the explanation of apparent large matter-anti-matter
%asymmetry through the leptogenesis. 
The determination of MH and the CP phase $\delta_{CP}$ will have profound significance 
not only within the neutrino physics, but in the larger field of high-energy physics.

\begin{figure}[]
\centering
\includegraphics[width=160mm]{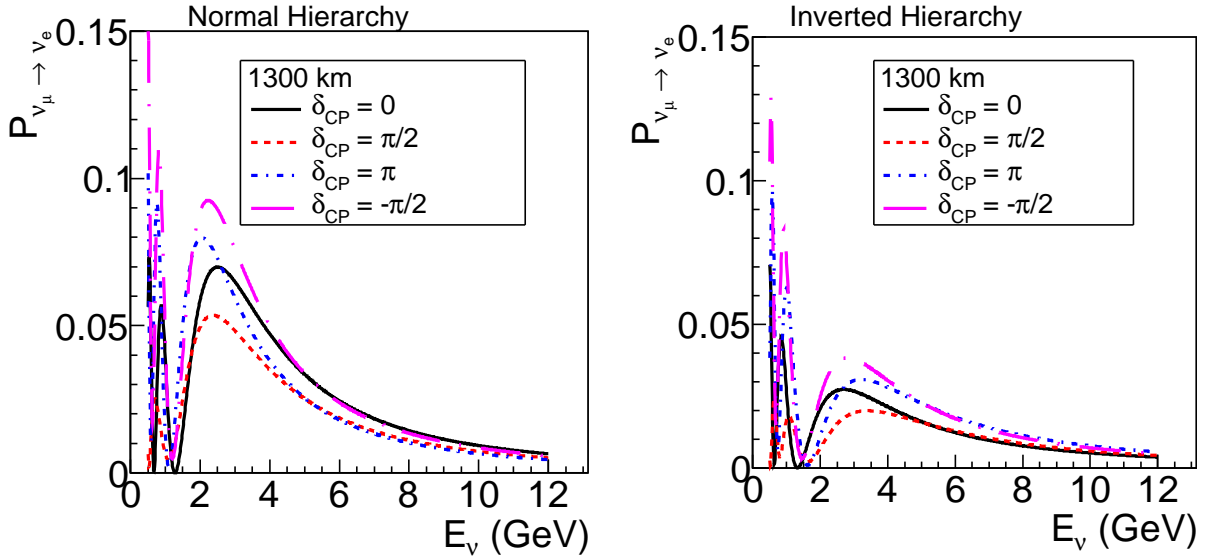}
\caption{ The $\nu_{\mu}$ to $\nu_{e}$ oscillation probabilities at a 1300 km baseline 
are shown for the normal (left) and inverted (right) mass hierarchy. The value of 
$\sin^{2}2\theta_{13}$ is assumed to be 0.092. On each panel, four curves corresponding 
to four values of $\delta_{CP}$ are shown.  
}
\label{fig:osc}
\end{figure}

\begin{figure}[]
\centering
\includegraphics[width=120mm]{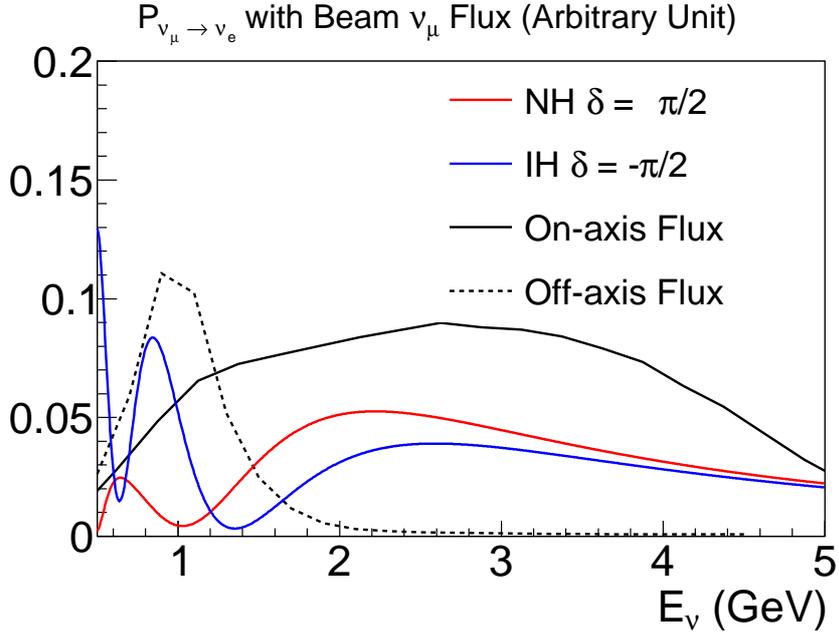}
\caption{The on-axis and 27 mrad off-axis neutrino beam energy profile 
at a baseline of 1300 km are shown. Together, the $\nu_{e}$ appearance 
probability for two cases: i) normal hierarchy with $\delta=\pi/2$ and ii)
inverted hierarchy with $\delta=-\pi/2$ are shown.  While the cancellation 
happens at the first oscillation maximum, the matter effect and the effect of $\delta_{CP}$ 
will result in a distinctive feature at the second maximum~\cite{low_energy_lbne}.
Furthermore, at second oscillation maximum, the off-axis beam provides a much higher 
flux than the on-axis beam. }
\label{fig:LBNE}
\end{figure}

The Long-Baseline Neutrino Experiment (LBNE)~\cite{lbne_cdr} is designed 
to determine the MH and $\delta_{CP}$ simultaneously. By placing an on-axis detector
at a baseline of 1300 km, LBNE will measure the (anti-)$\nu_{\mu}$ to (anti-)$\nu_{e}$ 
oscillation. As shown in Fig.~\ref{fig:osc}, the $\nu_{\mu}$ to $\nu_{e}$ 
oscillation probabilities are sensitive to both the MH (through the matter effect) and the value of
$\delta_{CP}$. Commonly, the peaks around 2.5 GeV and 0.8 GeV are 
referred to as the first and the second oscillation maximum, respectively. 
For the first oscillation maximum, the neutrino appearance probability is higher in the 
case of the normal hierarchy than in the case of the inverted hierarchy, regardless of 
the value of $\delta_{CP}$. With a wide-band (large energy coverage) beam, 
the LBNE on-axis detector will cover both the first and the second oscillation 
maximum (shown in Fig.~\ref{fig:LBNE}) with the emphasis on the first oscillation maximum. 

\begin{figure}[]
\centering
\includegraphics[width=120mm]{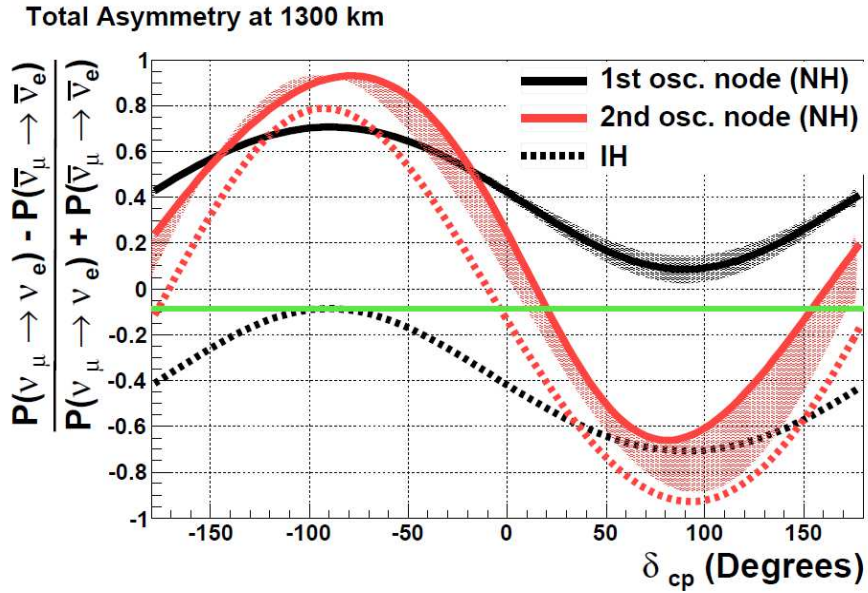}
\caption{The asymmetry between the $\nu_{e}$ and anti-$\nu_e$ appearance
at first and second oscillation maxima for both the normal and the inverted mass hierarchy.
}
\label{fig:mary}
\end{figure}

Due to the potential cancellation between the matter effect and the effect 
of $\delta_{CP}$ at the first oscillation maximum, the discovery potential of the MH strongly 
depends on the true value of $\delta_{CP}$.
%For example, the appearance probability of $\delta_{CP}=\pi/2$ with the normal 
%hierarchy is close to the appearance probability of $\delta_{CP}=-\pi/2$ with 
%the inverted hierarchy. Naturally, the ambiguity in MH weaken the potential
% to determine $\delta_{CP}$. 
For example, as seen in Fig.~\ref{fig:mary}, at the first oscillation maximum, the 
asymmetry in the neutrino-antineutrino appearance probability for $\delta_{CP}=\pi/2$ 
with the normal hierarchy is close to that for $\delta_{CP}=-\pi/2$ with the inverted hierarchy. 
On the other hand, for the second oscillation maximum, the size of the CP asymmetry is 
larger and the size of the matter effect is smaller, so the degeneracy between the 
two solutions is broken. Therefore, coverage of the second oscillation maximum is 
important for LBNE and an improved measurement of electron (anti-)neutrino appearance 
at the second oscillation maximum will enhance sensitivities to the MH and CP violation.
%On the other hand, as shown in 
%Fig.~\ref{fig:LBNE}, while the cancellation happens at the first maximum, 
%the matter effect and the effect of $\delta_{CP}$ will result 
%in a distinctive feature at the second maximum~\cite{low_energy_lbne}
%Therefore, the coverage of the second
%oscillation maximum is essential in LBNE, and an
%improved measurement at second oscillation maximum will enhance the 
%MH sensitivity. 

\begin{figure}[]
\centering
\includegraphics[width=160mm]{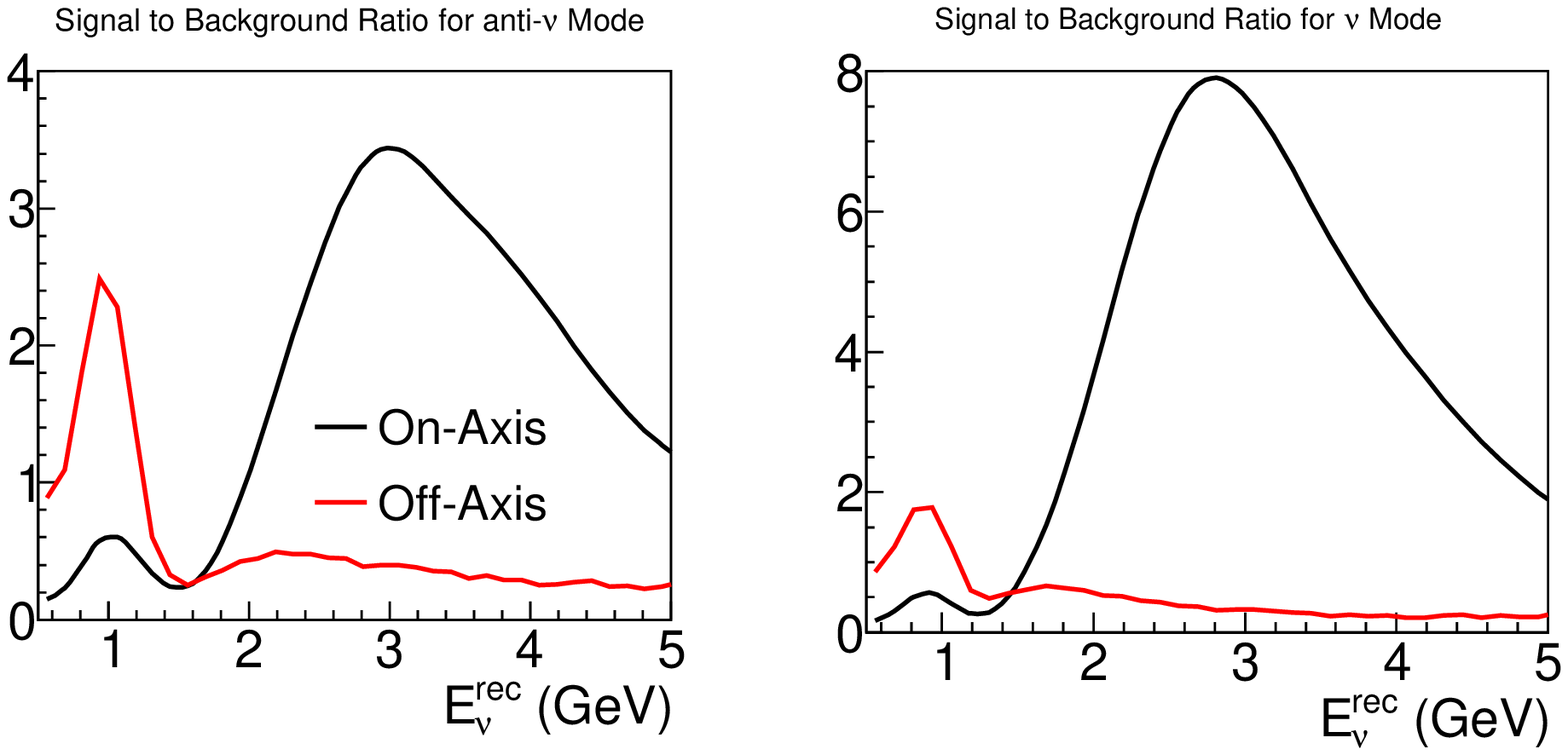}
\caption{ The signal-to-noise ratios for $\nu$ (right) and anti-$\nu$ (left) running 
are shown with respect to the reconstructed neutrino energy ($E_{\nu}^{rec}$). 
A 10 kt liquid argon time projection chamber is assumed to be the on-axis detector. 
A 10 kt water Cerenkov detector is assumed to be the off-axis detector. The 
signal-to-noise ratio at the first (second) oscillation is excellent for
the on-axis (off-axis) detector. 
}
\label{fig:sob}
\end{figure}

In this whitepaper, {\it we propose a second detector, a 10 kt water Cerenkov 
detector, at an off-axis 
location dedicated to the 
second oscillation maximum to enhance the MH discovery potential and the overall 
performance of the LBNE program.}

Due to the two-body kinematics of the pion decay, the off-axis neutrino flux 
is narrow in energy~\cite{off_axis1,off_axis2}. 
Fig.~\ref{fig:LBNE} shows the neutrino flux using the beam design described in 
Ref.~\cite{lbne_report} at 27 mrad off-axis.
The peak of the off-axis neutrino beam approximately matches the location of 
the second oscillation maximum. Comparing to the on-axis neutrino beam, the off-axis beam 
flux is much higher at the second oscillation maximum. The narrow-band 
beam also provides advantages in reducing backgrounds, which are 
generated by high-energy neutrinos, but misidentified as being 
low-energy neutrinos due to imperfect reconstructions. 
These include neutral-current  background and $\tau$-neutrino (oscillated from the 
$\mu$-neutrino) appearance background.
 Fig.~\ref{fig:sob} shows the expected 
signal-to-background ratio for the (anti-)$\nu_{\mu}$ running (NH and $\delta_{CP}$=0) 
with a 10 kt off-axis water Cerenkov (WC) detector in which the performance is assumed to 
be the same as Super-Kamiokande (SK2 performance~\cite{sk,wc_eff} assumed). 
The expected signal-to-background ratios
for the 10 kt LBNE on-axis liquid argon time projection chamber (LAr TPC) are also shown in Fig.~\ref{fig:sob} 
for comparison. The signal-to-background ratio is excellent at the first oscillation 
maximum for the on-axis LAr TPC detector, while the off-axis WC detector provides 
a cleaner signal at the second oscillation maximum.
In addition, it has been pointed out in Ref.~\cite{E889} that such a 
detector can work on surface to reduce cost with sufficient 
shielding.

\begin{figure}[]
\centering
\includegraphics[width=160mm]{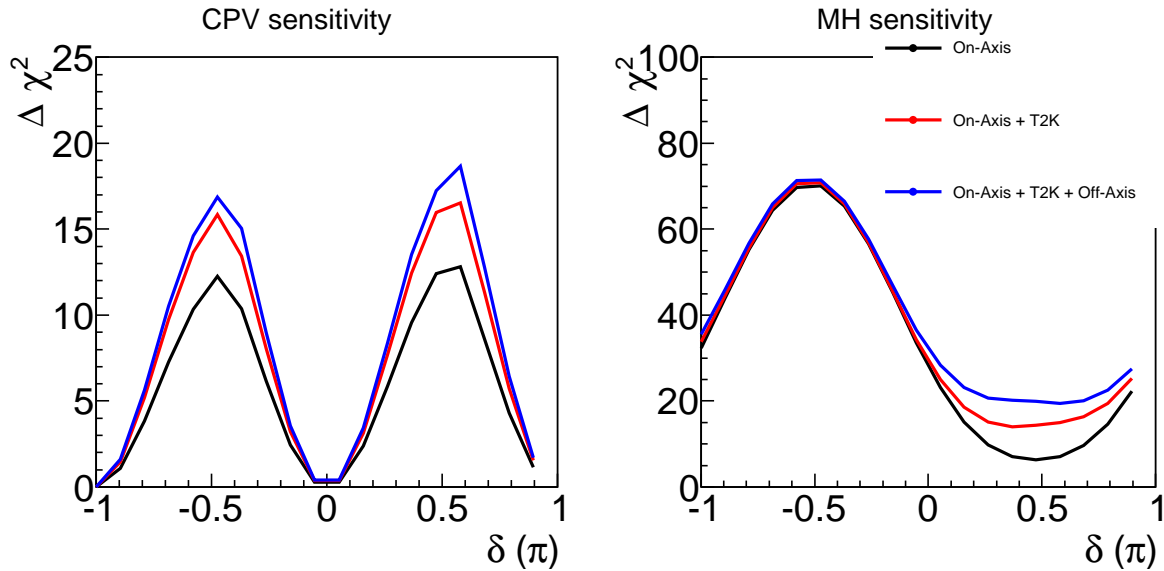}
\caption{ The physics sensitivity of LBNE10 to determine that the 
value of the CP-violating phase, $\delta_{CP}$, is not zero or $\pi$ 
(left) and to determine the neutrino mass hierarchy (right) are shown.
Three cases are compared: i) LBNE10 with
only the on-axis 10 kt LAr TPC, ii) combining the results of LBNE10 with 
those from the T2K experiment, and iii) combining the results of LBNE10 with T2K and 
a 10-kt water Cerenkov detector at a location 27-mrad off-axis from 
the LBNE beam line. The mass hierarchy sensitivity in the bad half 
of $\delta_{CP}$ is significantly improved with the second off-axis detector. 
}
\label{fig:sen}
\end{figure}

The physics sensitivities to the determination of the mass hierarchy and 
the discovery of CP violation with the additional 10-kt, off-axis water 
Cerenkov detector, are calculated in GLoBES~\cite{lbne_globes} using the 2010 
LBNE beam design  (Fig.~\ref{fig:sen}). The input oscillation parameters to
generate expected spectra are taken from Ref.~\cite{global_fit}.
%The physics sensitivities to MH and $\delta_{CP}$ with the additional 10 kt 
%water Cerenkov detector (27 mrad or 35 km off-axis at a distance of 1300 km) 
%are calculated in GLoBES with the 2010 LBNE flux and the 
%up-to-date neutrino oscillation parameters~\cite{global_fit} 
The statistical interpretation of MH sensitivity is explained in detail 
in Ref.~\cite{mh_stat}. The off-axis detector will considerably 
improve the combined sensitivity of LBNE10 and
 T2K~\cite{nu_fact,JHF-Kamioka, double-detector} in MH, with slight 
improvement in sensitivity to the CP violation. At the worst possible 
$\delta_{CP}$ values, the increment in the MH sensitivity 
($\Delta \chi^2$) with the second 10 kt off-axis WC detector
is equivalent to that of an additional 10 kt on-axis LAr TPC.
Since the WC detector is much cheaper than the LAr TPC with 
the same target mass, adding the second off-axis WC detector 
is a more efficient way to enhance MH sensitivity at 
the bad half of $\delta_{CP}$.
In addition, the MH sensitivity can be further 
enhanced with a narrower off-axis neutrino beam~\cite{details} without 
increasing the total flux intensity or a larger target mass~\footnote{
The newly proposed 100 kt water Cerenkov detector in mine
pits (CHIPS)~\cite{chips} could be re-deployed in 
Belle Fourche Reservoir, a lake located at about 27 mrad off-axis (1300 km) of 
the LBNE beam.}.

\begin{figure}[]
\end{figure}

In summary, a second detector at an off-axis location focusing on the second 
oscillation maximum can enhance the ability of LBNE10 to determine the mass 
hierarchy and strengthen its potential to discover CP violation in the leptonic sector.
%In summary, a second detector at an off-axis location focusing on the second 
%oscillation maximum can enhance the MH discovery potential in LBNE10 and
%strengthen the measurement of CP phase $\delta_{CP}$ in the leptonic sector. 

\end{document}